\documentclass[10pt,conference]{IEEEtran}
\IEEEoverridecommandlockouts
\usepackage{amsmath,amssymb,amsfonts}%编辑数学公式的宏包
\usepackage{textcomp}
\usepackage{xcolor}    %颜色
\usepackage{wasysym} 
\usepackage{amsthm}  %编辑数学定理和证明过程的宏包
\usepackage{graphicx,times,cite, epsfig, array, mathrsfs}  %插入图片的宏包
\usepackage{array}
\usepackage{multirow}  %复杂表格的宏包
\usepackage[caption=false,font=normalsize]{subfig}
\usepackage{textcomp}
\usepackage{stfloats}
\usepackage{url}
\usepackage{verbatim}
\usepackage{algorithm}
\usepackage[noend]{algorithmic}
\usepackage{caption}
\usepackage{threeparttable}
\usepackage{cases}
\def\BibTeX{{\rm B\kern-.05em{\sc i\kern-.025em b}\kern-.08em
    T\kern-.1667em\lower.7ex\hbox{E}\kern-.125emX}}

\newtheorem{theory}{Theorem}
\newtheorem{lemma}{Lemma}
\newtheorem{remark}{Remark}

\begin{document}

\title{
%Opportunistic Probing and  User Scheduling for Cache-Aided Hybrid Satellite-Terrestrial Networks
Joint Probing and Scheduling for Cache-Aided Hybrid Satellite-Terrestrial Networks
\vspace{-0.2cm}
}
%==============================================================================================================================
\author{ %\vspace{-0.1cm}
Zhou~Zhang\IEEEauthorrefmark{1}, 
Yizhu~Wang\IEEEauthorrefmark{1},
Saman~Atapattu\IEEEauthorrefmark{2} 
and Sumei Sun\IEEEauthorrefmark{3}
% \vspace{-0.3cm}
\\
 \IEEEauthorblockA{
% \IEEEauthorrefmark{1}National Innovation Institute of Defense Technology, Beijing, China.
 \IEEEauthorrefmark{1} Hunan University, Changsha, Hunan, China. \\
% \IEEEauthorrefmark{2}School of Engineering, RMIT University, Melbourne, Victoria, Australia. \\
 \IEEEauthorrefmark{2}Department of Electrical and Electronic Engineering, School of Engineering, RMIT University, Melbourne,  Australia. \\
\IEEEauthorrefmark{3}Institute for Infocomm Research, Agency for Science, Technology and Research, Singapore \\
\IEEEauthorblockA{Email:
\IEEEauthorrefmark{1}\{zt.sy1986, wangyizhuj\}@163.com;\,
\IEEEauthorrefmark{2}saman.atapattu@rmit.edu.au;\,
\IEEEauthorrefmark{3}sunsm@i2r.a-star.edu.sg 
% \vspace{-0.2cm}
}
}
% \thanks{This work was supported in part by the Australian Research Council (ARC) Discovery Project under Grant DP220103281 and Future Fellowship under Grant FT210100728;
% and in part by the National Natural Science Foundation of China under Grant 62171456.}
\vspace{-1cm}
}

\maketitle

\begin{abstract}
Caching is crucial in hybrid satellite-terrestrial networks to reduce latency, optimize throughput, and improve data availability by storing frequently accessed content closer to users, especially in bandwidth-limited satellite systems, requiring strategic Medium Access Control (MAC) layer. This paper addresses throughput optimization in satellite-terrestrial integrated networks through opportunistic cooperative caching. We propose a joint probing and scheduling strategy to enhance content retrieval efficiency. The strategy leverages the LEO satellite to probe satellite-to-ground links and cache states of multiple cooperative terrestrial stations, enabling dynamic user scheduling for content delivery. Using an optimal stopping theoretic approach with two levels of incomplete information, we make real-time decisions on satellite-terrestrial hybrid links and caching probing. Our threshold-based strategy optimizes probing and scheduling, significantly improving average system throughput by exploiting cooperative caching, satellite-terrestrial link transmission, and time diversity from dynamic user requests. Simulation results validate the effectiveness and practicality of the proposed strategies.
% \com{better to mention what mathematical/theoretical techniques used!}
\end{abstract}

\vspace{-1mm}
\section{Introduction} \label{s:intro}
\vspace{-0.1cm}
% Low Earth Orbit (LEO) satellite networks provide essential infrastructure for wide-reaching coverage and content delivery, seamlessly integrating with terrestrial networks to meet diverse traffic demands, from oceanic monitoring to live streaming. However, satellite links can be obstructed by terrain features like buildings and mountains, making hybrid satellite-terrestrial frameworks essential. In such systems, terrestrial relays play a key role in ensuring consistent performance through cooperative transmission~\cite{An2015,Zhu2023}. Applications that require high-speed data retrieval, such as video streaming, real-time analytics, and IoT monitoring, increasingly rely on satellite communications infrastructure and benefit greatly from caching strategies. In these hybrid systems, caching popular content at terrestrial nodes minimizes delays by allowing frequently accessed data to be stored locally, reducing reliance on satellite bandwidth and network congestion~\cite{An2019}. Without caching, content delivery from Internet gateways through satellite and ground stations extends service times, consuming limited satellite bandwidth and impacting the system's ability to support latency-sensitive applications. To optimize spectrum efficiency in such dynamic networks, combining caching with opportunistic user scheduling is crucial. This paper investigates joint caching and scheduling strategies in hybrid satellite-terrestrial networks to boost overall performance~\cite{zhao2023wcl}.

Low Earth Orbit (LEO) satellite networks enable wide-area, high-capacity communication for services such as maritime surveillance, real-time media, and IoT. However, terrain blockages degrade link reliability, motivating hybrid satellite-terrestrial architectures with terrestrial relays~\cite{Zhu2023}. To reduce latency and alleviate satellite bandwidth constraints, edge caching at relay nodes is essential~\cite{An2019,Guo2024}. Without caching, delay-sensitive services suffer from high end-to-end latency. While caching combined with opportunistic scheduling can enhance spectral efficiency~\cite{Zhao2022,Jiaran2023,zhang2025IoTJ}, this integration remains underexplored, making it the core focus of this work.

% Low Earth Orbit (LEO) satellite networks offer wide-area coverage and high-capacity content delivery, complementing terrestrial infrastructure to support diverse services such as maritime monitoring, real-time media and IoT connectivity. However, terrain-induced blockages can impair satellite links, motivating hybrid satellite-terrestrial architectures that incorporate terrestrial relays to enhance link reliability~\cite{An2015,Zhu2023}. Emerging applications, e.g., video streaming, real-time analytics, and IoT monitoring increasingly depend on satellite communications, where edge {\it caching} at terrestrial nodes reduces latency, conserves satellite bandwidth and mitigates network congestion~\cite{An2019}. Without caching, satellite-ground content delivery suffers from high latency, limiting delay-sensitive services. While  
%  caching with opportunistic user scheduling can improve spectral efficiency~\cite{Zhao2022,Jiaran2023}, this integration remains largely unexplored. Thus it is the key focus of our work. %~\cite{zhao2023wcl}.

%Zxuan_twc2024
\vspace{-1mm}
\subsection{Related Work} 
\vspace{-1mm}

Recent research on joint satellite–terrestrial user scheduling has demonstrated significant improvements in coverage and spectral efficiency by leveraging both satellite and terrestrial links~\cite{Zhu2023,An2019,Guo2024,Han2022,zhao2023wcl,zhang2023distributedjournal,zhang2024Journal,Ji2020,Zhao2021,Yting2021,Zhao2022,Jiaran2023,zhang2025IoTJ}.
%\com{check if i separate the correctly?}
In {\it non-caching scenarios}, 
%\cite{An2015} showed that a terrestrial relay improves downlink performance in terms of ergodic capacity, outage, diversity, and coding gain. For uplink, 
\cite{Zhu2023} proposed a low-complexity scheduling algorithm that maximizes sum rate using a terrestrial relay. Channel utilization was further enhanced in \cite{Han2022} with a multi-relay NOMA scheme and a three-stage selection strategy for full diversity. 
Using a distributed leaning approach, \cite{Zhao2021} proposed novel terrestrial relays selection and access scheme for massive IoT devices uplink transmission
to improve the system throughput. To address user dynamics, \cite{Yting2021} proposed a dynamic channel allocation strategy that prioritizes high-priority IoT devices while ensuring access for lower-priority users. Additionally, \cite{zhao2023wcl,zhang2024Journal} introduced an active user probing and random access scheme for dynamic resource allocation to maximize throughput. 
For a multi-cell satellite-terrestrial network, \cite{Ji2020} developed a terrestrial station aided data offloading mechanism to maximize downlink capacity and energy efficiency. 
In {\it caching-aided access}, \cite{An2019,Guo2024} proposed relay-assisted transmission with cached content to enhance reliability and reduce outage probability. 
In multi-satellite networks, \cite{Zhao2022} introduced a cooperative video caching algorithm for adjacent satellites, optimizing cache placement using a multi-agent deep deterministic policy gradient approach. For satellite IoT networks, \cite{Jiaran2023} developed a cooperative cache placement scheme that uses wireless caching at both satellite and terrestrial stations to improve caching service time, with a satellite-weighted time-varying graph for content caching along the data delivery path.

\vspace{-1mm}
\subsection{Problem Statement and Contributions}
\vspace{-1mm}
Although user scheduling \cite{Zhu2023, zhang2025IoTJ,Han2022, zhao2023wcl, Yting2021, zhang2024Journal, Zhao2021} and cache-aided content delivery \cite{An2019,Guo2024,Zhao2022, Jiaran2023} have been independently studied in satellite–terrestrial networks, their joint optimization remains largely unexplored. Existing works on cache-enabled hybrid architectures primarily target static users and focus on reducing outage or latency, often overlooking dynamic user requests, time-varying channel state information (CSI), cache probing overhead, and CSI  acquisition costs. Efficient scheduling in such networks must address the coexistence of satellite and terrestrial caches, random demands and varying channels, requiring adaptive strategies that exploit cached content, reduce overhead, and optimize delivery across heterogeneous links. 

Motivated by the need to enhance throughput under dynamic network conditions, this paper addresses the {\it cache-aided hybrid satellite–terrestrial joint probing and scheduling (HSTJPS)} problem at the Medium Access Control (MAC) layer. The main contributions are: 
i) {\it Opportunistic Probing and Scheduling:} An adaptive, overhead-aware framework is proposed to jointly manage user requests, channel dynamics, and cache states for efficient hybrid access. 
ii) {\it Hierarchical Optimal Stopping:} A novel two-level optimal stopping model is developed to enable real-time decision-making for joint probing and scheduling. 
iii) {\it Threshold-Based HSTJPS Policy:} An optimal threshold policy is derived, striking a balance between link efficiency and signaling overhead, achieving superior throughput over baseline schemes.

\vspace{-1mm}
%\com{As diverse contents could compete for scarce caching resources, resulting in network inefficiency and content delays.}
\section{System Model}\label{s:system_model}
\vspace{-1mm}
%\section{System model and }
%\vspace{-1mm}
%\subsection{System Model}\label{s:system_model}
 \begin{figure}[t!]
\begin{center}
\includegraphics[scale=.35]{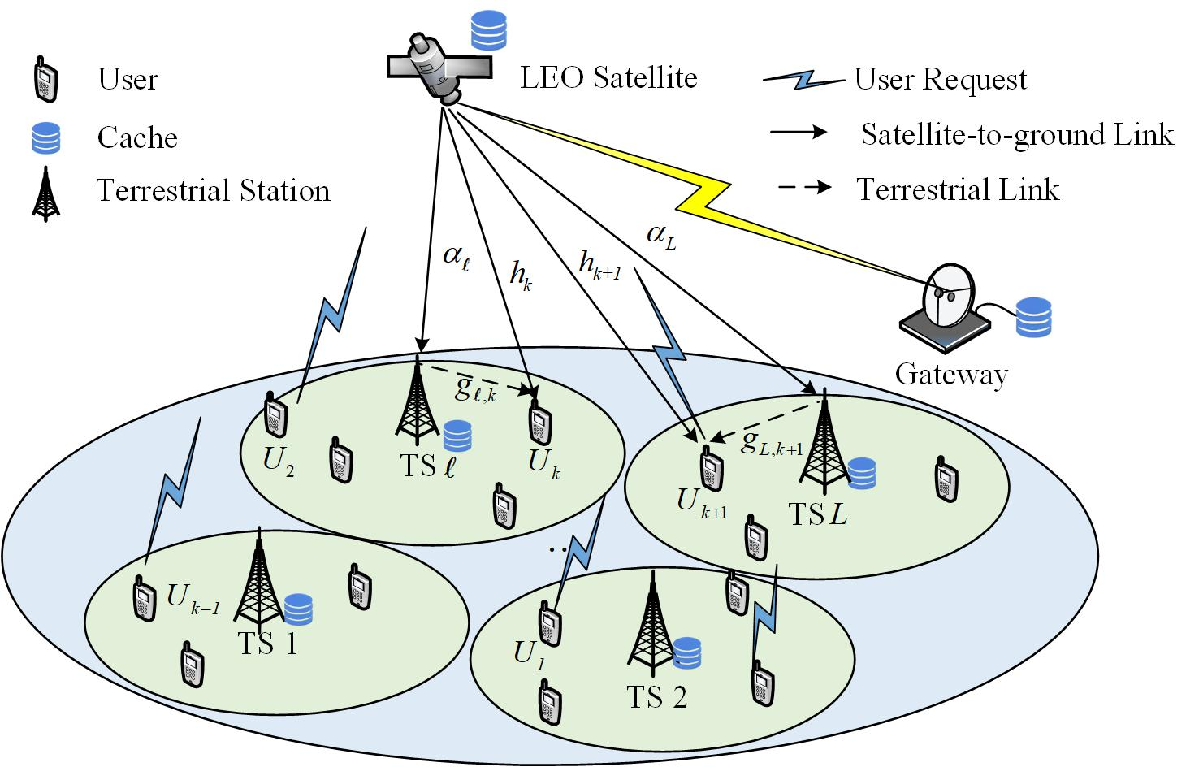} % system_model.eps
\caption{Illustration of a hybrid satellite-terrestrial network 
%\com{label as Terrestrial Station (TS), LEO Satellite, use diff colours for sat-ground and teresstrial links, make font size larger for all figures! }}
\vspace{-3mm}}\label{f:system_mod2}
\end{center}
\vspace{-5mm}
\end{figure}

We consider a hybrid satellite-terrestrial system where users $U_k, k \in \mathbb{N}$ request content files from a passing LEO satellite linked to a ground gateway through backhaul, as shown in Fig.~\ref{f:system_mod2}. User request times $t_k$ follow a Poisson process with rate $\lambda_s$, implying inter-arrival times $\tau_k = t_k - t_{k-1}$ are independent, exponentially distributed with mean $\tau_s = 1/\lambda_s$~\cite{Park2023}. 
The content files requested are denoted by $\{f_1, \ldots, f_I\}$, with file sizes $\{b_1, \ldots, b_I\}$ in units of bits. File popularity follows a Zipf distribution, $p_i = i^{-\zeta}/(\sum_{u=1}^I u^{-\zeta})$, where $\zeta$ is the skewness parameter~\cite{An2019}. 
With coverage difference between the satellite beam and terrestrial networks, the satellite coverage area is divided into $L$ cells, each served by a terrestrial station (TS) at the cell center. Let $C_s$ and $C_t$ represent the cache capacities of the satellite and TSs in units of bits. Using a probabilistic caching model, the caching probability of file $f_i$ at the satellite and TSs is $p_i^s$ and $p_i^t$, respectively, with $\sum_{i=1}^I p_i^s b_i = C_s$ and $\sum_{i=1}^I p_i^t b_i = C_t$. %~\cite{ZhangS2020}.

\subsection{Wireless Link Model}\label{sub:mac_channel}
% For cell $\ell$ with radius $R$, we consider a 2D geometric model with TS $\ell$ fixed at the cell center point $z_\ell$, respectively.
% Users are located in all $L$ cells with equal probability. Let $d_{0}$ denote the distance from users to the satellite.
% For satellite-to-ground links between the satellite and ${U}_k$, channel gain $h_{k}$ 
% follows a Shadowed-Rician (SR) distribution, i.e. $|h_{k}|\sim SR(\Omega_1,b_s,m_s)$ where $\Omega_1$ denotes the average power of the LoS component, $m_s$ is the fading order, and $2b_s$ represents for the average power of the scattered component~{\cite{An2019}}.
% Similarly, for links between the satellite and TS $\ell$, channel coefficients $\alpha_{k}$ follow the SR distribution with $|\alpha_{k}|\sim SR(\Omega_2,b_t,m_t)$.
% Moreover, in each cell users are uniformly distributed with probability density $1/\pi R^2$.
% For user $U_k$ at point $z=(z_X,z_Y)$ located within cell $\ell$, let $d_{k}$ denote the inter-distance $||z||_2$ from the user to TS $\ell$.
% Moreover, terrestrial wireless links between a user and its located cell's TS follow the Rayleigh fading, where channel coefficients
% ${g}_{\ell,k}=d_{k}^{-2\beta_0}\cdot \overline{g}_{\ell,k}$ % respectively, 
% with $\overline{g}_{\ell,k}\sim\exp(1)$ and terrestrial link path loss exponent $\beta_0$.  
% For actual file transmission, we consider finitely variable rates denoted as $R_m, m=0,1,...,M$ with $R_0<R_1<\ldots<R_M$, with corresponding SNR demodulation thresholds $\gamma_{m}, m=1,...,M$ and $\gamma_{0}=0$.

For each cell $\ell$ with radius $R$, we model a 2D geometry with the TS located at the cell center  $z_\ell$. Users are uniformly distributed across all $L$ cells. The satellite-to-user distance is $d_0$. 
The antenna gains of the satellite, TSs and users are 
denoted as $G_s$, $G_t$ and $G_u$, respectively.
The satellite-to-user links and satellite-to-TS links follow land mobile satellite (LMS) fading model,
with channel gains for $U_k$ and for TS $\ell$ denoted as $h_k$ and $\alpha_{\ell}$, respectively.
%the channel gain $h_k$ for user $U_k$ follows a Shadowed-Rician (SR) distribution, $|h_k| \sim SR(\Omega_1, b_s, m_s)$, where $\Omega_1$ is the average LoS component power, $m_s$ is the fading order, and $2b_s$ is the average power of the scattered component~{\cite{An2019}}. Similarly, for satellite-to-TS links, channel coefficients $\alpha_k$ follow $|\alpha_k| \sim SR(\Omega_2, b_t, m_t)$.
Within cell, users are uniformly distributed with probability density $1/(\pi R^2)$. For user $U_k$ at position $z = (z_X, z_Y)$ in cell $\ell$, let $d_k = ||z||_2$ denote the distance to TS $\ell$. Terrestrial links between a user and its cell's TS experience Rayleigh fading, with channel coefficients given by $|{g}_{\ell,k}|^2  \sim \exp(1)$. The path loss exponent is $\beta_0$.
For file transmission, we use discrete rates $R_m$, $m = 0, 1, \ldots, M$, with $R_0 < R_1 < \ldots < R_M$, and associated SNR thresholds $\gamma_m$, $m = 1, \ldots, M$, with $\gamma_0 = 0$.
 \vspace{-2mm}
\subsection{Opportunistic User Scheduling Scheme}\label{sub:mac_scheme}
% We propose the cooperative satellite-terrestrial probing and user scheduling scheme, an opportunistic approach that leverages the optimal stopping decision concept for efficient user scheduling of randomly arrived file retrieval requests as follows.

% At the start of a file retrieval time frame ($t_0=0$), the satellite awaits requests from its users. The process unfolds as the following steps:
% i) after a random time interval $\tau_k$, User $U_k$ transmits its request for file $f_i$ with pilots to the satellite with location information. The satellite then acquires the file size $Q_k$, its caching state $\beta_{s,k}\in\{0,1\}$, instantaneous sat-user link channel gain $h_{k}$, and TS $\omega_{k}\in\{1,...,L\}$ where user $U_k$ is located.
% ii) the satellite calculates the instantaneous reward $\Lambda_1$ for Sat-User delivery and the expected reward $\Lambda_2$ for TS cooperative transmission based on random variables $(Q_k,\beta_{s,k},h_{k})$. It then compares the achievable rewards $\Lambda_1$ and $\Lambda_2$ with the threshold $\Lambda_3$ and chooses one of three options. The design parameters are discussed in Section~\ref{s:optimal-DOCA-stratg}.
We propose a cooperative satellite-terrestrial joint probing and scheduling scheme, which uses an opportunistic approach with optimal stopping for efficiently scheduling randomly arriving file retrieval requests. The process operates as follows:
At the start of each file retrieval time frame ($t_0 = 0$), the satellite awaits user requests. The steps are:
%The opportunistic scheduling process follows these steps:
i) After a random time interval $\tau_k$, user $U_k$ sends a request for file $f_i$, including pilots and location data, to the satellite. The satellite then obtains the file size $B_k$, caching state $\beta_{s,k} \in \{0, 1\}$ ($\beta_{s,k}=1$ denotes the file is cached and verse vice.), the instantaneous Sat-User link channel gain $h_k$, and the user's TS location $\omega_k \in \{1, \dots, L\}$.
ii) The satellite calculates the instantaneous reward $\Lambda_1$ for direct satellite delivery and the expected reward $\Lambda_2$ for cooperative transmission via the TS, based on $(B_k, \beta_{s,k}, h_k)$. The satellite then compares $\Lambda_1$ and $\Lambda_2$ against a threshold $\Lambda_3$ and selects one of three actions accordingly. Design parameters are discussed in Section~\ref{s:optimal-DOCA-stratg}.
\begin{enumerate}
    \item {\bf {Direct Satellite Delivery}}: If $\Lambda_1 \ge \max\{\Lambda_2, \Lambda_3\}$, the satellite schedules user $U_k$ for direct file retrieval during $T_{k,1} = B_k/R_s(k) + T_1 + \mathbb{I}[\beta_{s,k} = 0] T_2$, using the transmission mode in Section~\ref{sub:relay_scheme1}. Here, $R_s(k)$ is the Sat-User transmission rate, $T_1$ is the satellite-ground link latency, and $T_2$ is the time needed to fetch the file from the gateway if not cached locally. 
    % \footnote{Since the satellite-gateway link is typically LoS, its channel gain is stable and can be considered fixed over long durations.}.
    \item {\bf Wait}: If $\max\{\Lambda_1, \Lambda_2\} < \Lambda_3$, the satellite drops $U_k$ and waits for the next user $U_{k+1}$.
    \item {\bf TS Probe}: If $\Lambda_2 > \max\{\Lambda_1, \Lambda_3\}$, the satellite instructs TS $\omega_k$ to probe its Sat-TS link gain $\alpha_{\omega_k,k}$, TS-User link gain $g_{\omega_k, k}$, and cache status $\beta_{r,\omega_k} \in \{0, 1\}$ within a duration $\tau_p$
    % \footnote{Probing time $\tau_p$ includes the latency $T_1$, terrestrial channel estimation and TS processing time.}
    ($\beta_{r,\omega_k}=1$ denotes the file is cached at the TS and verse vice.). TS then calculates the reward $\Gamma_1$ for assisted delivery and compares it to a threshold $\Gamma_2$ (details in Section~\ref{s:optimal-DOCA-stratg}). The TS selects the optimal delivery mode based on cache status $\beta_{r,\omega_k}$ and channel gains $\{\alpha_{\omega_k,k}, g_{\omega_k, k}\}$: If $\beta_{r,\omega_k} = 1$, TS delivers the file directly to $U_k$; If $\beta_{r,\omega_k} = 0$, the satellite delivers to $U_k$ relaying through TS $\omega_k$.
    \begin{enumerate}
        \item {\bf Assisted Delivery}: If $\Gamma_1 \ge \Gamma_2$, TS schedules user $U_k$ for delivery during $T_{k,2} = \mathbb{I}[\beta_{r,\omega_k} = 1] B_k/R_{r,1}(k) + \mathbb{I}[\beta_{r,\omega_k} = 0] \left(B_k/R_{r,2}(k) + T_1\right)$, as described in Section~\ref{sub:relay_scheme2}.
        \item {\bf Wait}: If $\Gamma_1 < \Gamma_2$, the satellite drops $U_k$ and waits for the next user.
    \end{enumerate}
\end{enumerate}
A new time frame starts upon scheduling a user.

 \vspace{-2mm}
\subsection{File Transmission Modes}\label{sub:relay_scheme}
 \vspace{-2mm}
This section describes the optimal transmission modes and latency for both direct sat-user and assisted delivery.
 %\vspace{-1mm}
\subsubsection{Sat-User transmission}\label{sub:relay_scheme1}
For direct delivery to user $U_k$, the transmission latency is $T_{k,1} = B_k / R_s(k) + T_1 + \mathbb{I}[\beta_{s,k} = 0] T_2$. The achievable rate $R_s(k)$ is given by
 \vspace{-2mm}
\begin{align*}
R_s(k) = \sum_{m=1}^M \mathbb{I}[\gamma_m \le \gamma_{s,k} < \gamma_{m+1}] R_m,
\end{align*}
where the received SNR is $\gamma_{s,k} = \overline{\gamma}_s|h_k|^2$, average SNR $\overline{\gamma}_s=P_{t,s}G_sG_td_0^{-2}/N_0$,
and $P_{t,s}$ is the satellite's power.
%\vspace{-1mm}
\subsubsection{Assisted delivery transmission}\label{sub:relay_scheme2}
For cache-aided transmission from TS $\omega_k$ to user $U_k$, the rate based on caching state $\beta_{r,\omega_k} = 1$ and TS-User link gain $g_{\omega_k,k}$ is
\vspace{-1mm}
\begin{align*}
R_{r,1}(k) = \sum\limits_{m=1}^M \mathbb{I}[\gamma_m \le \gamma_{1,k} < \gamma_{m+1}] R_m,
\end{align*}
%\vspace{-1mm}
where the received SNR is $\gamma_{1,k} = \overline{\gamma}_u|g_{\omega_k,k}|^2$, the average SNR at the user 
is $\overline{\gamma}_u =P_{t,r}G_tG_ud^{-2\beta_0}/N_0$, and $P_{t,r}$ is the TS transmit power. 
For relaying by TS $\omega_k$, with no cached file ($\beta_{r,\omega_k} = 0$) and channel gains $(\alpha_{\omega_k,k}, g_{\omega_k,k})$, the rate is
\vspace{-1mm}
\begin{align*}
R_{r,2}(k) = \sum_{m=1}^M \mathbb{I}[\gamma_m \le \gamma_{2,k} < \gamma_{m+1}] R_m,
\end{align*}
%\vspace{-1mm}
where $\gamma_{2,k} = \overline{\gamma}_s |h_k|^2 + \frac{\overline{\gamma}_t |\alpha_{\omega_k,k}|^2 \overline{\gamma}_u|g_{\omega_k,k}|^2}{\overline{\gamma}_t  |\alpha_{\omega_k,k}|^2 + \overline{\gamma}_u |g_{\omega_k,k}|^2}$, and the average SNR at the TS is $\overline{\gamma}_t =P_{t,s}G_sG_td_0^{-2}/N_0$, respectively.

%The first term represents the latency by {\it mode I delivery}, where the BS delivers to $\mathcal{D}_k$ via the best decode-and-forward (DF) relay among probed nodes in ${\mathcal L}_k$. The second term represents the latency by {\it mode II delivery}, where the BS and nodes that have stored the file cooperatively deliver to $\mathcal{D}_k$.
%\subsubsection{TS relaying transmission}
%The first term represents the latency by {\it mode I delivery}, where the BS delivers to $\mathcal{D}_k$ via the best decode-and-forward (DF) relay among probed nodes in ${\mathcal L}_k$. The second term represents the latency by {\it mode II delivery}, where the BS and nodes that have stored the file cooperatively deliver to $\mathcal{D}_k$.

\section{Problem Formulation}\label{s:homo}
We use an optimal stopping approach with two levels of incomplete information to formulate an optimization problem for the proposed scheme, aiming to maximize average system throughput in opportunistic scheduling.

\subsection{Preliminaries of Optimal Stopping Approach}\label{s:theory}
The optimal stopping approach leverages observed information to reach target outcomes, solving for the optimal observation path and timing. The problem is defined by
\begin{itemize}
    \item {\bf Sequence of Random Variable Vectors}: We have a sequence $\{\mathbf{X}_1, \mathbf{Y}_1, \dots, \mathbf{X}_n, \mathbf{Y}_n, \dots\}$, where $\mathbf{X}_n = \{X_{n,1}, \dots, X_{n,M_1}\}$ are i.i.d. with distribution $\mathbf{X}_n \sim \mathbf{X}_0$, and $\mathbf{Y}_n = \{Y_{n,1}, \dots, Y_{n,M_2}\}$ are i.i.d. with distribution $\mathbf{Y}_n \sim \mathbf{Y}_0$.
    \item {\bf Feasible Observation Path (OP)}: Defined as $\mathbb{A} = {(a_1, \dots, a_{2n}) : n \in \mathbb{N}}$, where $a_{2n-1} = 1$ means $\mathbf{X}_n$ is always observed, and $a_{2n} \in \{0, 1\}$ indicates whether $\mathbf{Y}_n$ is observed.
    \item {\bf Sequence of Reward Functions}: This is a sequence $\big\{z_1(\mathbf{x}_{1},\mathbf{y}_{1}(a_2))-c_0-a_2c_1,\ldots,z_{2n-1}(\mathbf{x}_{1},\mathbf{y}_{1}(a_2),\mathbf{x}_{2},...,\mathbf{x}_{n})-n c_0-\sum\limits_{i=1}^{n-1} a_{2i}c_1,z_{2n}(\mathbf{x}_{1},\mathbf{y}_{1}(a_2),\mathbf{x}_{2},...,\mathbf{x}_{n},\mathbf{Y}_{n}(a_{2n}))-n c_0-\sum\limits_{i=1}^n a_{2i}c_1\big\}$, where rewards are based on observed RVs in the feasible OP, with costs $c_0$ and $c_1$ for observing each $\mathbf{X}_n$ and $\mathbf{Y}_n$, respectively.
\end{itemize}
In the optimization problem, the decision-maker selects random variables to observe at each step and decides whether to {\it stop} or {\it continue}, aiming to maximize the expected reward. The goal is to determine the optimal OP and stopping time, denoted as $N=(a_1,a_2, \ldots)$, to achieve $Z^* {=} \sup\limits_{N}\mathbb{E}[Z_N]$, where $Z_N=z_N\big(\mathbf{X}_{1},\mathbf{Y}_{1}(a_2),\ldots,\mathbf{X}_{\lceil |N|/2 \rceil}, \mathbf{Y}({a_{|N|}})\big)$ is the random reward upon stopping.

To solve the optimization problem, for OP $\mathbf{a} = (a_1, \dots, a_{2n-1})$ at time step $(2n-1)$, 
{we define the reward expectation function 
\begin{equation}\label{equ:general}
    {\Gamma}_{(\mathbf{a},1)}(\mathbf{X}_n){=}\mathbb{E}\big[\max\{Z_{(\mathbf{a},1)},Z^*-T_c(2n)\}|\mathbf{X}_n\big],
\end{equation}
%$$,
where $T_c(2n)$ is the total cost for previous observations. This represents the conditional expected maximum reward from observing additional information $\mathbf{Y}_n$ after observing $\mathbf{X}_n$.}
Theorem~\ref{th:optimal_rule} provides the optimal stopping rule based on these expectation functions.

% \begin{theory}\label{th:optimal_rule} 
% An expectation function threshold (EFT) based rule, denoted as $N_1$, is defined as follows: Starting from 
% $n=0$, at time step $(2n-1)$, the rule makes following decisions:
% \begin{enumerate}
% \item if $Z_{\mathbf{a}}\ge \max\big\{U_0-T_c(2n-1),{\Gamma}_{(\mathbf{a},1)}(\mathbf{X}_n)\big\}$, to stop with $N_1=\mathbf{a}$,
% where $T_c(2n-1)=n c_0+c_1\sum_{j=1}^{n-1}\mathbb{I}_{[a_{2j}>0]}$ denotes the total time cost for previous observations.
% \item if $\max\big\{Z_{\mathbf{a}},{\Gamma}_{(\mathbf{a},1)}(\mathbf{X}_n)\big\}<U_0$, to skip observing $\mathbf{Y}_n$ and proceed to time step $(2n+1)$;
% \item otherwise, to observe $\mathbf{Y}_n$;
% \end{enumerate}
% At time step $2n$ after observing $\mathbf{Y}_n$, the rule has following decisions: if $Z_{\mathbf{a}}\ge U_0-T_c(2n)$, to stop with $N_1=\mathbf{a}$; otherwise, proceed to time step $(2n+1)$.

% Moreover, if $\mathbb{E}\big[\sup\limits_{{\mathbf a}\in{\mathbb{A}}}Z_{{\mathbf a}}\big]<+\infty$, %the existence condition is met, 
% the EFT-based rule $N_1$ is optimal, i.e. $N^*=N_1$, achieving $U_0=\sup\limits_{N} \mathbb{E}[Z_N]$. 
% %The maximal reward $U_0$ is uniquely determined by 
% The maximal expected reward $U_0$ is determined by % the equation:
% \begin{align}\label{e:optimality_equation}
%    U_0=&\mathbb{E}\big[\{Z_{(1)},U_0,{\Gamma}_{(1,1)}(\mathbf{X}_0)\}\big]-c_0.
% \end{align}    
% \end{theory}
% \begin{IEEEproof}
% It can be derived by Theorem 1 in \cite{Fenoy2017}.
% \end{IEEEproof}

\begin{theory}\label{th:optimal_rule} 
The optimal stopping rule, denoted as $N_1$, is based on an expectation function threshold (EFT). Starting at $n = 0$, at time step $(2n-1)$, the decisions are:
\begin{enumerate}
\item If $Z_{\mathbf{a}} \ge \max\big\{Z^* - T_c(2n-1), {\Gamma}_{(\mathbf{a},1)}(\mathbf{X}_n)\big\}$, stop with $N_1 = \mathbf{a}$, where $T_c(2n-1) = n c_0 + c_1 \sum_{j=1}^{n-1} \mathbb{I}{[a_{2j} > 0]}$ is the total cost.
\item If $\max\big\{Z_{\mathbf{a}}, {\Gamma}_{(\mathbf{a},1)}(\mathbf{X}_n)\big\} < Z^*$, skip observing $\mathbf{Y}_n$ and proceed to time step $(2n+1)$.
\item Otherwise, observe $\mathbf{Y}_n$.
\end{enumerate}
At time step $2n$, after observing $\mathbf{Y}_n$, if $Z_{\mathbf{a}} \ge Z^* - T_c(2n)$, stop with $N_1 = \mathbf{a}$; otherwise, proceed to time step $(2n+1)$.

If $\mathbb{E}\big[\sup_{\mathbf{a} \in \mathbb{A}} Z_{\mathbf{a}}\big] < +\infty$, the EFT-based rule $N_1$ is optimal, achieving $Z^*$. The maximal expected reward $Z^*$ is given by
\begin{align}\label{e:optimality_equation}
  Z^*= \mathbb{E}\big[\{Z_{(1)}, Z^*, {\Gamma}_{(1,1)}(\mathbf{X}_0)\}\big] - c_0.
\end{align}    
\end{theory}
\begin{IEEEproof}
It can be derived from Theorem 1 in \cite{Fenoy2017}.
\end{IEEEproof}

\subsection{Optimal Stopping Approach Based Problem Formulation}\label{s:homo2}

We maximize system throughput by optimizing the observation process of file requests, caching states, and channel conditions and stopping time under the opportunistic hybrid satellite-terrestrial delivery framework, using the optimal stopping approach outlined in Section \ref{s:theory}.

\subsubsection{Observation Process}
% Following the HSTPUS scheme in Section \ref{sub:mac_scheme}, the satellite and terrestrial stations act as the decision-maker, starting from the first user request ($k=1$). The observation involves acquiring relevant random variables for each user $U_k$, which include:
% \begin{itemize}
%     \item The satellite observes $\mathbf{X}_k=\big\{Q_k,\beta_{s,k}, h_{k},\omega(k),d(k)\big\}$ after a random interval $\tau_k$, capturing RVs information $\mathbf{X}_k$ and makes a decision at time step $(2k\!-\!1)$.
%     \item If the decision is to observe more, TS $\omega(k)$ further observes $\mathbf{Y}_k= \big\{\beta_{t,\omega(k)}(k), \alpha_k,g_{\omega(k),k}\big\}$.
% \end{itemize}
% For the OP, each request logs two numbers for every step: $a_{2k\!-\!1} = 1$, ensuring $\mathbf{X}_k$ is observed, and $a_{2k}$, indicating wether $\mathbf{Y}_k$ is observed. To optimize rewards, decisions are based on prior observations up to a decision to stop. Correspondingly, the OP is denoted as $\mathbf{a} = (a_1,a_2, \ldots,a_{2k})$.

\begin{itemize} 
\item The satellite observes $\mathbf{X}_k=\big\{B_k,\beta_{s,k}, h_{k},\omega_k,d_k\big\}$ after a random interval $\tau_k$ and makes a decision at time step $(2k\!-\!1)$. 
\item If the decision is to observe more, TS $\omega_k$ observes $\mathbf{Y}_k= \big\{\beta_{r,\omega_k}(k), \alpha_{\omega_k,k},g_{\omega_k,k}\big\}$
and makes a decision at time step $2k$. 
\end{itemize} 
For the OP, each request logs two numbers: $a_{2k\!-\!1} = 1$, ensuring $\mathbf{X}_k$ is observed, and $a_{2k}$, indicating whether $\mathbf{Y}_k$ is observed. The OP is denoted as $\mathbf{a} = (a_1, \dots, a_{2k})$.
%, and feasible observation paths are defined by the set $\mathbb{A} = \{(a_1,a_2, \ldots,a_n) : n \in \mathbb{N}, a_{2k-1}=1, a_{2k} \in \{0, 1\}, \forall k \leq n/2 \}$.

\subsubsection{Reward and Cost Functions}
The reward for stopping at OP $\mathbf{a}$, denoted $V_{\mathbf{a}} = B_k$, is the requested file size. The associated time cost $T_{\mathbf{a}}$ is the total duration from Obs. 1 to Obs. $|\mathbf{a}|$ plus latency. At time step $|\mathbf{a}| = 2k-1$,  
$T_{\mathbf{a}} = \sum_{j=1}^k \tau_j + \sum_{j=1}^{k-1} \mathbb{I}{[a_{2j}>0]} \tau_p + T_{k,1} $;  
and at time step $|\mathbf{a}| = 2k$,  
$T_{\mathbf{a}} = \sum_{j=1}^k \tau_j + \sum_{j=1}^k \mathbb{I}{[a_{2j}>0]} \tau_p + T_{k,2}$.

\subsubsection{Optimization Goal}
Let $N$ denote the sequential OP upon a {\it stop}. Here, $V_N$ represents the reward function and $T_N$ the time cost at each stop by strategy $N$. In the HSTJPS scheme for one time frame, the system throughput, defined as the ratio of the average reward to the average time cost, is $\mathbb{E}[V_N]/\mathbb{E}[T_N]$ [bits/s]. The goal is to find an optimal strategy $N^*$ that maximizes the average system throughput $\eta^*$:
% We denote $N$ as the sequential OP upon {\it stop}. Thus, $Y_N$ means the reward function upon a {\it stop} decision by strategy $N$, and $T_N$ means the time cost spent for each stop by strategy $N$. While it denotes an SPD rule, we use it to represent the decisions of the JCPUS strategy for one time content delivery. When strategy $N$ is performed repeatedly, the system throughput, which is the ratio of the average reward to the average time cost, can be expressed as $\mathbb{E}[Y_N]/\mathbb{E}[T_N]$ with unit {[bits/s]}. Accordingly, our goal is to find an optimal strategy $N^*$ achieving the maximal average system throughput $\eta^*$, which are given as  
\begin{equation}\label{equ:stat_opt}
	N^*=\arg\sup\limits_{N}\frac{\mathbb{E}[V_N]}{\mathbb{E}[T_N]}~~\text { and } ~~\eta^*=\frac{\mathbb{E}[V_{N^*}]}{\mathbb{E}[T_{N^*}]}.
\end{equation}

% \section{Optimal HSTPUS Strategy}\label{s:optimal-DOCA-stratg}
% We derive the optimal HSTPUS strategy $N^*$ maximizing average system throughput based on the EFT-based SPD rule.
% \vspace{-2mm}
% \subsection{Equivalent Ratio Optimization Problem}\label{sub:equal_transfer}
% To optimize the throughput ratio, $\sup\limits_{N} \mathbb{E}[V_N]/\mathbb{E}[T_N]$, we link it to a price-based objective. By introducing $\eta$ as the price on time cost, we define utility functions $Z_{\mathbf{a}}(\eta) = V_{\mathbf{a}} - \eta T_{\mathbf{a}}$ and $Z_N(\eta) = V_N - \eta T_N$. For a given $\eta > 0$, the rule achieving $\sup\limits_{N} \mathbb{E}[Z_N(\eta)]$ is denoted as $N(\eta)$, with an optimal rule represented by $N^*(\eta)$, which can be expressed as 
% \begin{equation} 
% 	N^*(\eta)=\arg\sup\limits_{N}Z_N(\eta)=\arg\sup\limits_{N}\{V_{N}-\eta T_{N}\}.
% \end{equation}
% The strategy $N^*(\eta^*)$ serves as the optimal $N^*$ for the ratio optimization problem, where $\eta^*$ is the unique value satisfying $\sup\limits_{N} \mathbb{E}[Z_N(\eta^*)] = 0$, and is given by $\eta^* = \sup\limits_{N} \mathbb{E}[V_N] / \mathbb{E}[T_N]$.

\section{Optimal HSTJPS Strategy}\label{s:optimal-DOCA-stratg}  
We now derive the optimal HSTJPS strategy $N^*$. 
% that maximizes average system throughput using the EFT-based SPD rule.
\vspace{-0.1cm}
\subsection{Equivalent Ratio Optimization Problem}\label{sub:equal_transfer}  
To optimize the throughput ratio $\sup\limits_{N} \mathbb{E}[V_N] / \mathbb{E}[T_N]$, we introduce a price-based objective with $\eta$ as the price on time cost. The utility functions are defined as $Z_{\mathbf{a}}(\eta) = V_{\mathbf{a}} - \eta T_{\mathbf{a}}$ and $Z_N(\eta) = V_N - \eta T_N$. For a given $\eta > 0$, the rule achieving $\sup\limits_{N} \mathbb{E}[Z_N(\eta)]$ is denoted $N(\eta)$, with the optimal rule $N^*(\eta)$: % given by  
\[
N^*(\eta) = \arg\sup\limits_{N} \{ V_{N} - \eta T_{N} \}.
\]  
The strategy $N^*(\eta^*)$ is the optimal solution for the ratio optimization, where $\eta^*$ satisfies $\sup\limits_{N} \mathbb{E}[Z_N(\eta^*)] = 0$, and is given by $\eta^* = \sup\limits_{N} \mathbb{E}[V_N] / \mathbb{E}[T_N]$.

%%%%%%%%%%% equation for Th 2

%%%%%%%%%%%%%%%
%, and based on which an optimal strategy $N^*$ is derived.
%maximizing the average system throughput $\sup\limits_{N}\mathbb{E}[Y_N]/\mathbb{E}[T_N]$.
%In this section, we find 
%\vspace{-1mm}
%\begin{remark}
%Being our main conclusion, Theorem \ref{th:optimal_rule2} provides a bridge between the optimal SPD strategy and DOS/SP strategy. The strategy from this theorem achieves the maximal average reward $\sup\limits_{N}\mathbb{E}[Z_N]$. Nevertheless, threshold-related functions $U_0$ and $\{M_j(\cdot)\}_{j=1,\ldots,L}$ requires further derivation. In the light of Lemma \ref{l:equ_problem_trans}, by replacing price $\lambda$ by maximal average system throughput $\lambda^*$ such that $U_0(\lambda^*)\!=\!0$, an optimal DOS/SP strategy is obtained.
%%Accordingly, in the sequel further analysis is made with respect to refinement of threshold functions in Theorem \ref{th:optimal_rule2}.
%\end{remark}
\vspace{-0.1cm}
\subsection{Optimal HSTJPS Strategy}
%We define reward function $M_\ell\big(q,h_s,\beta,d,L,\eta\big)$ as the expected reward if TS $\ell$ where user $U_k$ is located in the cell is probed, expressed as $M_\ell\big(h_s,\beta,q,\eta\big):=\mathbb{E}\big[\max\big\{q-\eta T_{k,2},0\big\}\big|h_k=h_s,\beta_{s,k}=\beta,Q_k=q,d_k=d,L=\ell\big]-\eta\tau_p$.
\vspace{-0.1cm}
Based on the expectation function $\Gamma_{(\mathbf{a},1)}(\mathbf{X}_n)$ in (\ref{equ:general}), we define the reward function as 
\begin{align}
     &\Omega(b_i,\beta,h_s,d,\eta,Z^*\big):=
\mathbb{E}\big[\max\big\{b_i-\eta T_{k,2},Z^*\big\}\big||h_k|^2\!=
\!h_s,\nonumber\\&\quad\quad
\beta_{s,k}=\beta,B_k=b_i,d_k=d,\omega_k=\ell\big]-\eta\tau_p.
\end{align}
It represents the maximal average reward if the located TS is probed for user $U_k$. 
By probability analysis under channel fading models, we have following results.

\begin{lemma}
The cumulative distribution function (CDF) of the received SNR for TS relaying transmission from the satellite to user $U_k$, conditioned on the Sat-User link channel gain $|h_k|^2 = h_s$ and TS-User distance $d_k= d$, is given by 
%==========
\begin{align*}
    &F_{\gamma_{2,k}}(x,d|h_s)  =1\!-\!e^{\left(\frac{x-\overline{\gamma}_{s}h_s}{\overline{\gamma}_u}\right)}\Bigg(1\!-\!\frac{\Omega_t(x-\overline{\gamma}_{s}h_s)(2b_t m_t)^{m_t}}{2b_t\overline{\gamma}_{t}(2b_t m_t+\Omega_t)^{m_t}} \nonumber\\
    & \times{}_1F_1\left(m_t,2,\frac{\Omega_t^2 (x-\overline{\gamma}_{s}h_s)}{2\overline{\gamma}_{t}b_t(2b_t m_t+\Omega_t)}\right) \!+\! \frac{\Omega_t^2}{8b_t^2}\left(\!\frac{2b_t m_t}{2b_t m_t+\Omega_t}\!\right)^{m_t} \nonumber\\
    & \times\left(\frac{x-\overline{\gamma}_{s}h_s }{\overline{\gamma}_{t}}\right)^2 
    % \nonumber\\
    % & \quad \times 
    {}_2 F_2\left(2,m_t;3,1;\frac{\Omega_t^2 (x-\overline{\gamma}_{s}h_s) } {2\overline{\gamma}_{t}b_t(2b_t m_t+\Omega_t)}\right)\Bigg)\label{equ:cdf_gamma3}
\end{align*}
%=========
% \( F_{\gamma_{2,k}}(x, d | h_s) \) as expressed in (\ref{equ:cdf_gamma3}),
where ${}_1 F_1(a,b,z)$ and ${}_2 F_2(a_1,a_2;b_1,b_2;z)$ are hypergeometric functions. Further, $(\Omega_s,m_s,2b_s)$ denotes the average LoS component power, the fading order, and the average power of the scattered component
of the Sat-User links, and $(\Omega_t,m_t,2b_t)$ denotes the  parameters of the Sat-TS links.

An expression of the reward function at \( Z^* = 0 \) is given as
\vspace{-0.1cm}
\begin{align}
      &\Omega\big(b_i,\beta,h_s,d,\eta,0\big)\!=\!
    p_{i}^t\!\!\sum\limits_{\substack{m=1,\\R_{m}\ge \eta}}^M\!\! \!\Big(\!\big(\!F_{\gamma_{1,k}}(R_{m+1},d)\!-\!F_{\gamma_{1,k}}(R_{m},d)\!\big)    \nonumber\\&
    \quad(b_i\!-\!\eta b_i/R_m)\Big)+
(1\!-\!p_{i}^t)\! \!\!
\sum\limits_{\substack{m=1,\\R_{m}\ge \frac{\eta}{b_i\!-\!\eta T_1}}}^M\!\!\!\Big( (b_i\!-\!\eta T_1\!-\!\eta  b_i/R_m)
    \nonumber
    \\&\quad\big(F_{\gamma_{2,k}}(R_{m+1},d|h_s)-F_{\gamma_{2,k}}(R_{m},d|h_s)\big)\Big) -\eta\tau_p  
\end{align}
where $F_{\gamma_{1,k}}$ follows an exponential distribution. %
\end{lemma}

Moreover, we derive the optimal strategy as follows. %in the following theorem.
\begin{theory}\label{th:optimalrule4}
The optimal HSTJPS strategy $N^*$ achieving the maximal average system throughput $\sup\limits_{N}\mathbb{E}[V_N]/\mathbb{E}[T_N]$
is as follows: starting from $k=1$,
for user $U_k$, 
the satellite obtains $\big\{B_k,\beta_{s,k},h_k,\omega_k,d_k\big\}$,
\begin{enumerate}
\item  if the immediate reward $ B_k-\eta^* T_{k,1} \ge \max\big\{\Omega(B_k,\beta_{s,k},|h_k|^2,d_k,\eta^*,0),0\big\}$, 
the satellite opts {\bf direct satellite delivery} by scheduling $U_k$; 
\item if the immediate reward 
$\max\big\{B_k-\eta^* T_{k,1},\Omega(B_k,\beta_{s,k},|h_k|^2,d_k,\eta^*,0)\big\}<0$,
the satellite opts {\bf wait};
\item otherwise, the satellite opts {\bf terrestrial station probe}.
Then, 
it lets TS $\omega_k$ further probe to obtain $\{\alpha_{\omega_k,k},g_{\omega_k,k},\beta_{r,\omega_k}\}$;
\begin{enumerate}
    \item if $\beta_{r,\omega_k}=1$ and $R_{r,1}(k)\ge \eta^*$, then {\bf assisted delivery} by cache-aided transmission from TS $\omega_k$ to $U_k$;
    \item if $\beta_{r,\omega_k}=0$ and $R_{r,2}(k)\ge \eta^*B_k/(B_k-\eta^*T_1)$, then {\bf assisted delivery} by relaying transmission from satellite to $U_k$ via TS $\omega_k$;
    \item otherwise, {\bf wait} until next user $U_{k+1}$.
\end{enumerate}
\end{enumerate}
The maximal throughput \( \eta^* \) is uniquely determined by \( \Lambda(\eta^*) = \eta^* \tau_s \), where \( \Lambda(\eta) \) is given as 
\vspace{-1mm}
\begin{align}
    \Lambda(\eta) := &\sum\limits_{i=1}^I p_i \Bigg(
    p_{i}^s \int\limits_{r=0}^R\int\limits_{h=0}^\infty\frac{2r}{R^2}
    \max\!\big\{b_i-\eta (T_{k,1}+T_1),0,
    \nonumber\\
    &\quad {\Omega}\big(b_i,1,h,r,\eta,0\big) \big\} f_{\gamma_{s,k}}(h)
    {\rm d}h {\rm d}r 
    \nonumber\\
    + &(1-p_{i}^s) \int\limits_{r=0}^R\int\limits_{h=0}^\infty\frac{2r}{R^2}
    \max\!\big\{b_i-\eta (T_{k,1}+T_1+ T_2),0,
    \nonumber\\
    &\quad {\Omega}\big(b_i,0,h,r,\eta,0\big) \big\} f_{\gamma_{s,k}}(h)
    {\rm d}h {\rm d}r \Bigg)
    \label{e:optimality_equation2}
    \\
    \text{with}~ &f_{\gamma_{s,k}}(x)=\frac{\Omega_s}{2b_s\overline{\gamma}_s}
    \big(\frac{2b_s m_s \overline{\gamma}_s}{2b_s m_s+\Omega_s}\big)^{m_s}
    \nonumber\\
    &\quad \frac{\Omega_s x}{2b_s\overline{\gamma}_s} {}_1F_1\Big(m_s,1,
    \frac{\Omega_s^2 \overline{\gamma}_s}{2 b_s \overline{\gamma}_s(2b_sm_s+\Omega_s)}\Big).
  \end{align} %in (\ref{e:optimality_equation2}).
      \vspace{-1mm}
%, and the channel gain \( h_k \) follows an exponential distribution with CDF \( F_s(x) \).
\end{theory}

\begin{IEEEproof}
Utilizing the equivalence transfer method outlined in Section~IV-A, we address the maximization problem of system throughput through a two-step approach:

\paragraph{Transfer the EFT-based rule $N^*(\eta)$ to the HSTJPS strategy} 

In accordance with the theoretical framework in Section~III-A, we derive the optimal stopping rule $N^*(\eta)$ as follows.
%to achieve the supremum of the expected value $\mathbb{E}[Z_N(\eta)]$ for a given price $\eta>0$ is addressed. This optimization process 
%we derived by proving the condition and transformation of the RET-based rule in 
%into the SPD rule $N^*(\eta)$.%a modified optimization problem.
By definitions of $V_{{\mathbf a}}$ and $T_{{\mathbf a}}$, we can prove %establish inequality 
$  \mathbb{E}\big[\sup\limits_{{\mathbf a}\in{\mathbb{A}}}Z_{{\mathbf a}}(\eta)\big]
\!\le \mathbb{E}\big[B_k\big]=\sum_{i=1}^I p_i b_i <+\infty$.

%Then, based on Theorem~1, we transform the EFT-based rule $N_1$ into optimal rule $N^*(\eta)$.

Utilizing the decision equivalence in Section~III-B, we transition the rule $N^*(\eta)$ in Theorem 1 to the strategy and analyze the optimal decision conditions following each observation.

First, we examine the case when $n=2k\!-\!1$. The satellite opts for {\it stop} when $Z_{\mathbf a}\ge  \max\big\{Z^*,{\Gamma}_{(\mathbf{a},1)}(\mathbf{X}_n)\big\}$; 
%otherwise, it prolongs the observation process by probing $J^*$ cooperative nodes, where $J^*=\min\big\{0\le \ell\le L:U_{({\mathbf a},\ell)}=V_{{\mathbf a}}\big\}$. 
Based on the result of Lemma 1, the condition for {\it stop} is rewritten as
\vspace{-1mm}
\begin{align*}
&B_k-\eta T_{k,1}
\ge \max\big\{{\Omega}\big(B_k,\beta_{s,k},|h_k|^2,d_k,\eta,Z^*\big),Z^*\big\}.
\end{align*}
If the decision is to observe $\mathbf{Y}_n$, the condition is rewritten as 
\vspace{-1mm}
\begin{align*}
%\overline{M}_{J^*}\!(h_k,\beta_{b,k},Q_k,\eta,U_0\!) \!=\! 
&{\Omega}(B_k,\beta_{s,k},|h_k|^2,d_k,\eta,Z^*)> 
\max\{B_k-\eta T_{k,1},Z^*\} 
\end{align*}
%This condition ensures that the maximal average reward when probing $J^*$ cooperative nodes is greater than or equal to a predetermined threshold $U_0$, guiding the optimal decision-making process.

Otherwise, the condition for skipping observing $\mathbf{Y}_k$ is similarly derived.

Then, we examine the case when $n\!=\!2k$. The satellite opts {\it stop} if $Z_{\mathbf a}\ge Z^*-\eta T_c(2k)$,
where $T_c(2k)=\sum_{j=1}^k \tau_j+\sum_{j=1}^{k}\mathbb{I}{[a_{2j}>0]}\tau_p$ denotes the total time consumed until the step.
Otherwise, the satellite opts for {\it wait}. Specifically, the condition for \textit{stop} is:
\vspace{-1mm}
\begin{align*}
Z_{\mathbf a}=&
B_k\!-\!\eta T_{k,2}-\eta T_c(2k)
\ge Z^*\!-\!\eta T_c(2k).
\vspace{-2mm}
\end{align*}

Combining aforementioned results, the optimal strategy $N^*(\eta)$ for $\sup\limits_{N}\mathbb{E}[Z_N(\eta)]$, is as follows:
Starting from $k=1$, for user $U_k$, the satellite obtains %receives information on 
$\mathbf{X}_k$,
%The decision-making process is as follows:
\begin{itemize}
	\item if the immediate reward $B_k-\eta T_{k,1}\ge \Omega\big(B_k,\beta_{s,k},|h_k|^2,d_k,\eta,Z^*\big),Z^*\Big\}$, the satellite opts \textbf{direct satellite delivery}.
	\item if $\max\big\{B_k-\eta T_{k,1},\Omega\big(B_k,\beta_{s,k},|h_k|^2,d_k,\eta,Z^*\big)\big\}<Z^*$, the satellite opts \textbf{wait}.
	\item otherwise, the satellite opts \textbf{terrestrial station probe}. Subsequently, by probing $\mathbf{Y}_k$,
	\begin{itemize}
		\item if the immediate reward $B_k-\eta T_{k,2}\ge Z^*$, then the TS opts \textbf{assisted delivery}.
		\item otherwise, the TS opts \textbf{wait} for the next user $U_{k+1}$.
	\end{itemize}
\end{itemize}
The maximal expected reward $Z^*$ is determined by % the equation:
\begin{align}\label{e:optimality_equation}
   Z^*=&\mathbb{E}\big[\max\big\{B_k-\eta T_{k,1},
    \nonumber\\& ~~~~~~Z^*,  
\Omega(B_k,\beta_{b,k},|h_k|^2,d_k,\eta,Z^*)\big\}\big]-\eta\tau_s.
\end{align}
%Consequently, the index $(k)$ can be omitted.
%\end{theory}

% \begin{figure*}[b!]
%   \hrule
% \vspace{-0.1cm}
%   \begin{align}
% \Lambda(\eta):=&\sum\limits_{i=1}^I p_i \Big(p_{i}^s \int\limits_{r=0}^R\int\limits_{h=0}^\infty\frac{2r}{R^2}
% \max\!\big\{b_i-\eta (T_{k,1}+T_1),0,
% {\Omega}\big(b_i,1,h,r,\eta,0\big) \big\} f_{\gamma_{s,k}}(h)  {\rm d}h {\rm d}r 
% \nonumber\\&+
% (1-p_{i}^s) \int\limits_{r=0}^R\int\limits_{h=0}^\infty\frac{2r}{R^2}
% \max\!\big\{b_i-\eta (T_{k,1}+T_1+ T_2),0,
% {\Omega}\big(b_i,0,h,r,\eta,0\big) \big\}   f_{\gamma_{s,k}}(h){\rm d}h {\rm d}r 
% \Big)
% \label{e:optimality_equation2}
% \\&
% \text{where}~ f_{\gamma_{s,k}}(x)=\frac{\Omega_s}{2b_s\overline{\gamma}_s}\big(\frac{2b_s m_s \overline{\gamma}_s}{2b_s m_s+\Omega_s}\big)^{m_s}\frac{\Omega_s x}{2b_s\overline{\gamma}_s} {}_1F_1\Big(m_s,1,\frac{\Omega_s^2 \overline{\gamma}_s}{2 b_s \overline{\gamma}_s(2b_sm_s+\Omega_s)}\Big).
% \nonumber   
% \end{align} 
% \vspace{-4mm}
% \end{figure*}
\vspace{-2mm}
\paragraph{Replace $\eta$ with $\eta^*$ and obtain optimal strategy $N^*$} 
By utilizing the equivalence transferring method and substituting $Z^*$ and $\eta$ with $0$ and $\eta^*$, the optimal strategy $N^*=N^*(\eta^*)$
is explicitly described in the theorem.
Leveraging the i.i.d. statistical property of RVs $\mathbf{X}_k$ and interval time $\tau_k$ for $k\in\mathbb{N}$, the right-hand side of (\ref{e:optimality_equation2}) 
remains constant $\forall k\ge 1$. 
\end{IEEEproof}
Based on Theorem~\ref{th:optimalrule4}, we have following remarks. 
\vspace{-1mm}
\begin{remark}
The maximal average throughput \( \eta^* \) is determined by network statistics and computed offline as the solution to \( \Lambda(\eta^*) = \eta^* \tau_s \). The reward function \( \Omega(B_k, \beta_{s,k}, |h_k|^2,d_k, \eta^*, 0) \), depending on file size \( B_k \), channel gain \( |h_k|^2 \), and caching state \( \beta_{s,k} \), can be efficiently calculated online using the analytical expression in (6). 
%\com{With the observed variables $(Q_k,\beta_{s,k},h_k)$, the optimal decision can be executed with a maximum complexity of $\mathcal{O}(L)$, enabling practical online implementation of the strategy.}
\end{remark}

\begin{figure}[h]
  % \vspace{-3mm}
	\begin{center}
		\includegraphics[scale=.36]{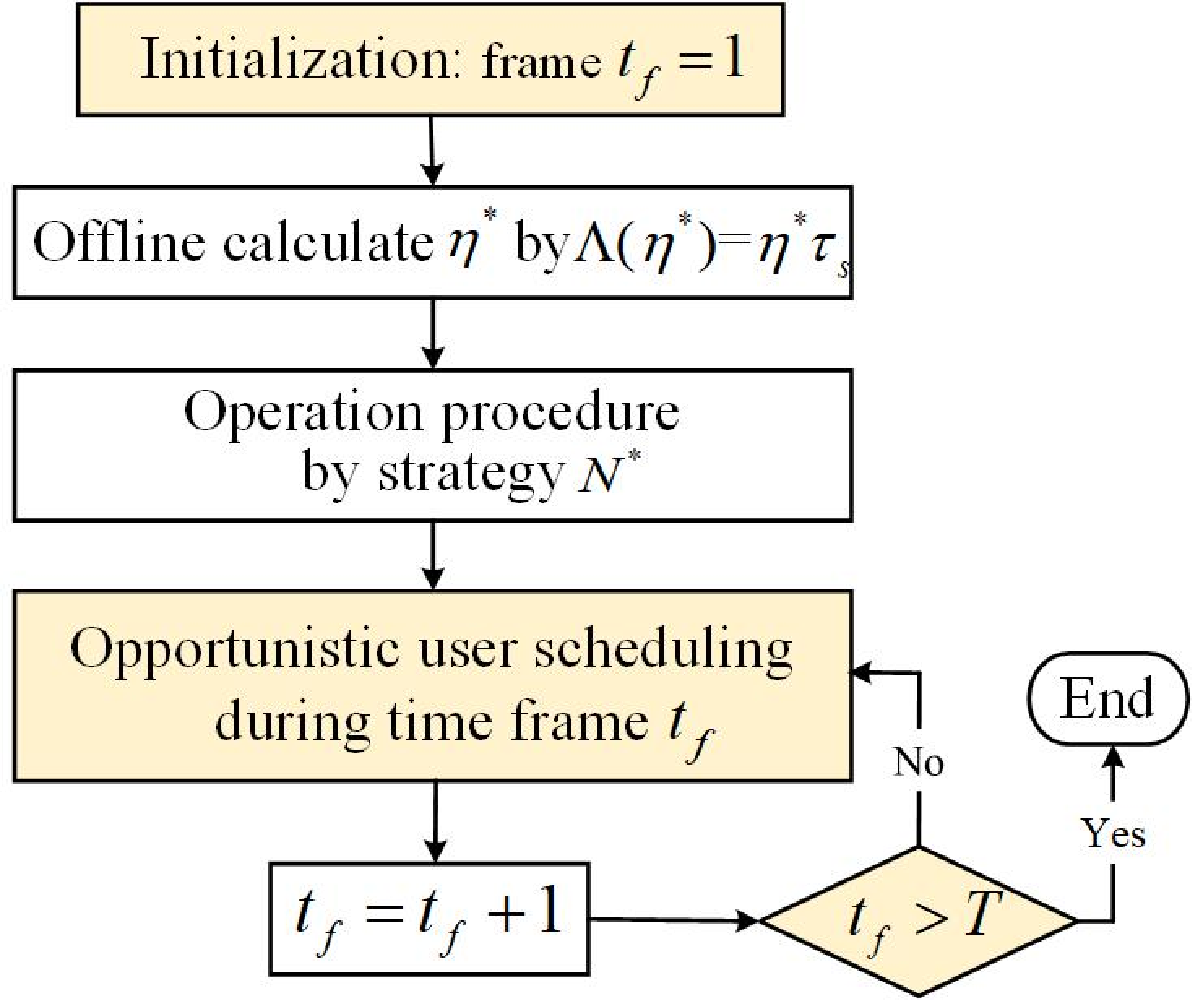}
		\caption{Flowchart for the strategy implementation.}\label{f:optimal_alg} %\com{do you have algo number to refer???}
	\end{center}
  \vspace{-5mm}
\end{figure}

\vspace{-2mm}

\begin{figure*}[h]
	\centering
	\subfloat[Throughput vs transmit power $P_{t,s}$.]{
		\label{fig:comparison1}
		\includegraphics[width=0.33\textwidth]{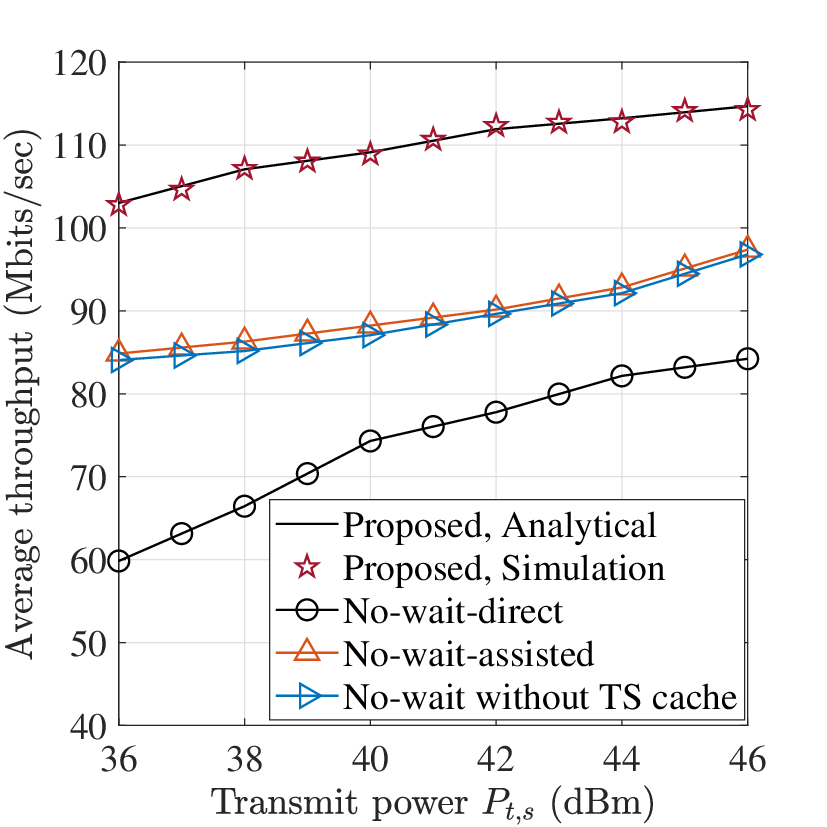}} 
  %\vspace{-1mm}
	\subfloat[Throughput vs transmit power $P_{t,r}$. ]{
		\label{fig:comparison2}
		\includegraphics[width=0.33\textwidth]{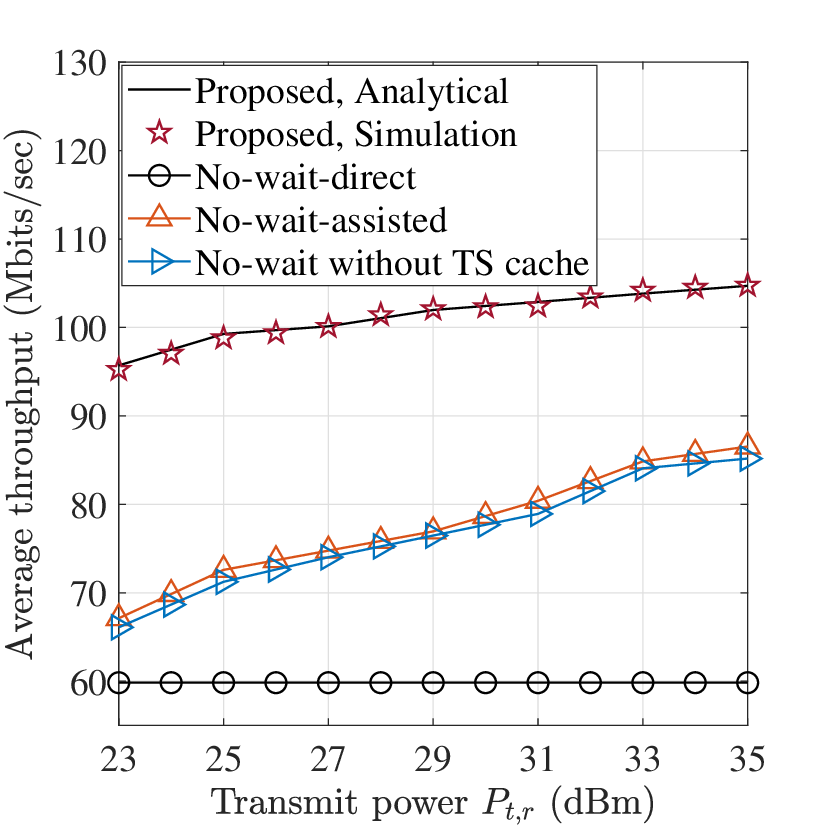}} 
  \subfloat[Throughput vs interval $\tau_s$.]{
		\label{fig:comparison3}
		\includegraphics[width=0.33\textwidth]{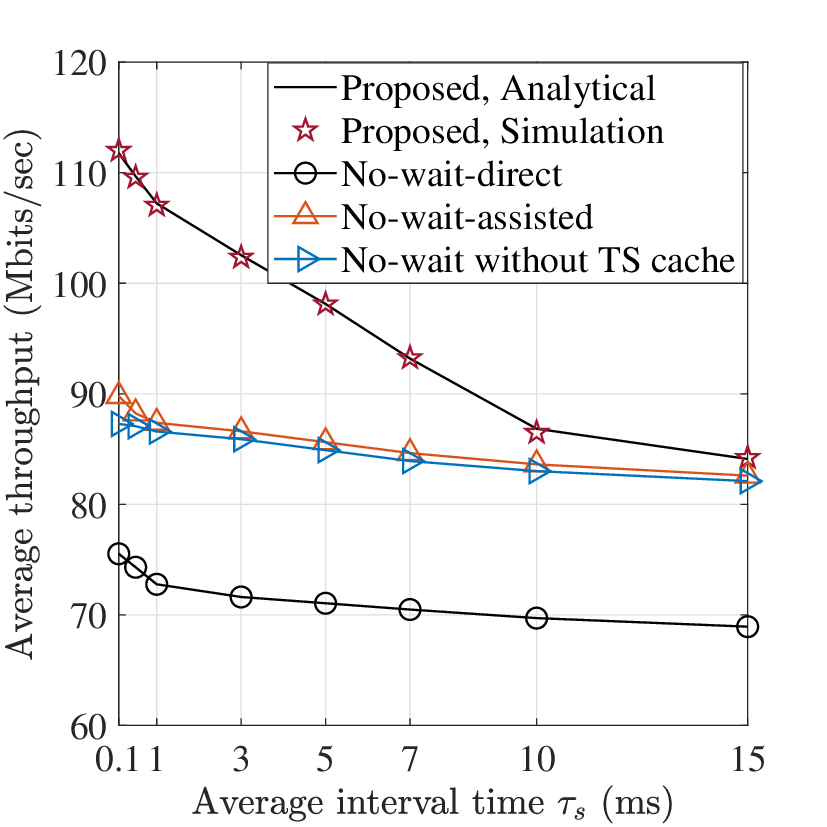}}
     \caption{Validation and average system throughput comparison with alternative strategies.
     \vspace{-1mm} }
	\label{fig:comparison}
 \vspace{-4mm}
\end{figure*}

\subsection{Strategy Implementation}\label{s:imp-DOCA-stratg}  \vspace{-1mm}
To maximize content retrieval efficiency and system throughput, the hybrid satellite-terrestrial system operates according to the flowchart in Fig.~\ref{f:optimal_alg}. At the initial file retrieval frame ($t_f = 1$), the system initializes the optimal strategy $N^*$ by computing the offline throughput $\eta^*$. Guided by the opportunistic user scheduling framework in Section~\ref{sub:mac_scheme}, the HSTJPS process executes iteratively over $T$ time frames. The strategy exploits unique hybrid network characteristics—spatial diversity from multiple terrestrial cells within a satellite beam and link heterogeneity %(Sat-Gateway, Sat-User, TS-User)-
to balance hybrid gains against latency, achieving higher throughput.

\vspace{-1mm}
\section{Numerical Results}\label{s:num}
\vspace{-1mm}
% This section carries out simulations to evaluate system performance.
We consider $L=7$ cells within the satellite coverage, with each cell radius at $R=1$\,km.
The LEO satellite is at altitude $600$~km, operating at $f_c=2$\,GHz with $20$\,MHz bandwidth.
The antenna gains of the satellite, TSs and the users are $25$\,dBi, $10$\,dBi and $0$\,dBi, respectively. 
For satellite-user link and satellite-TS link, the path-loss exponent is $2$, and the fading parameters are set as 
$\{m_s,b_s,\Omega_s\}=\{5,0.126,0.279\}=\{m_t,b_t,\Omega_t\}=\{5,0.126,0.279\}$~[3].
For terrestrial links, the path-loss exponent is $\beta_0=3.76$, and a reference path loss is $-40$\,dB at $1$\,m.
The noise power density is $N_0=-174$\,dBm/Hz. Moreover, for content files, the network manages $I=8$ content items with skewness $\zeta=1.5$ and file sizes $b_i=100$\,Mbits 
for $\forall i=1,...,I$.
The satellite cache holds $C_b=300$ Mbits with equal caching probabilities of $3/8$, and each TS stores $C_r=100$ Mbits with equal caching probabilities of $1/8$. 
The TS probing time is $\tau_p=2$\,ms, the transmission delay $T_1=2$\,ms, and the latency for the satellite fetching file from the gateway is $T_2=2$\,ms.
For multi-rate transmission, the transmission rates $[R_1,\dots,R_6]$ are set as $[43.3,57.8,86.7,115.6,130,144]$\,Mbps,
with corresponding demodulation thresholds $[\gamma_1,\dots,\gamma_6]$ are set as $[5.4,8.1,12.8,17.3,19.5,21.6]$\,dB.

Fig.~\ref{fig:comparison} compares average system throughput against satellite transmit power $P_{t,s}$, TS transmit power $P_{t,r}$, and request interval $\tau_s$, validating our theoretical results. It shows the match between analytical results from solving bellman equation $\Lambda(\eta^*) = \eta^* \tau_s$ and simulation outcomes from the HSTJPS strategy using Theorem~2, confirming the accuracy of our theory. Additionally, the figure benchmarks our strategy against three alternatives:  
i) \textit{No-wait-direct} strategy: The satellite only probes the direct link and its caching status, and schedules arrived user for connect delivery~\cite{An2019};  
ii) \textit{No-wait-assisted} strategy: The satellite schedules users promptly by probing both direct and terrestrial links and caching state of terrestrial stations, utilizing TS assisted delivery~\cite{An2019}; iii) \textit{No-wait without TS cache} strategy: The satellite schedules users promptly by probing both direct and terrestrial links of terrestrial stations, utilizing TS relaying delivery~\cite{Han2022}.

%\vspace{-0.1cm}
% Fig.~\ref{fig:comparison1} evaluates average system throughput versus satellite transmit power $P_{t,s}$ for $P_{t,r}=33$\,dBm and $\tau_s=0.5$\,ms. 
% As $P_{t,s}$ varies from $36$ to $46$~dBm, our strategy outperforms alternatives, achieving gains of at least 36.2\% over the no-wait-direct,
% and 18.4\% over the no-wait without TS cache, and 17.9\% over the no-wait-assisted strategies. Throughput increases with $P_{t,s}$, with our strategy maintaining superiority across all power levels.
% Fig.~\ref{fig:comparison2} shows average throughput versus TS transmit power $P_{t,r}$ for $P_{t,s}=36$\,dBm and $\tau_s=0.5$\,ms. Our strategy outperforms alternatives, with a 66.6\% gain over no-wait-direct, a 36.8\% over the no-wait without TS cache, and a 35.3\% advantage over no-wait-assisted strategies at $P_{t,r}=26$\,dBm. 
% {Throughput increases with larger transmit power, highlighting the benefits of time diversity.}
% In addition, Fig.~\ref{fig:comparison3} shows average throughput versus users request interval $\tau_s$ at $P_{t,s}=40$\,dBm and $P_{t,r}=33$\,dBm. Our strategy outperforms alternatives, with a 32.2\% energy efficiency gain over no-wait-direct and a 10.1\% throughput advantage over no-wait-assisted strategies at $\tau_s=7$\,ms. Throughput decreases as $\tau_s$ increases, with our strategy’s advantage diminishing for $\tau_s \geq 15$\,ms due to reduced time diversity benefits.
Fig.~\ref{fig:comparison1} plots average throughput versus satellite transmit power $P_{t,s}$ with $P_{t,r}=33$\,dBm and $\tau_s=0.5$\,ms. 
For $P_{t,s}\in[36,46]$\,dBm, the proposed strategy consistently outperforms baselines, achieving gains of $36.2\%$, $18.4\%$, and $17.9\%$ over the no-wait-direct, no-wait (without TS cache), and no-wait-assisted schemes, respectively. Throughput grows with $P_{t,s}$, with our scheme superior across all power levels.  
Fig.~\ref{fig:comparison2} shows throughput versus TS transmit power $P_{t,r}$ for $P_{t,s}=36$\,dBm and $\tau_s=0.5$\,ms. At $P_{t,r}=26$\,dBm, the proposed strategy yields $66.6\%$, $36.8\%$, and $35.3\%$ improvements over the no-wait-direct, no-wait (without TS cache), and no-wait-assisted strategies, respectively. Throughput increases with $P_{t,r}$, highlighting the benefit of time diversity.  
Fig.~\ref{fig:comparison3} illustrates throughput versus request interval $\tau_s$ at $P_{t,s}=40$\,dBm and $P_{t,r}=33$\,dBm. At $\tau_s=7$\,ms, the proposed strategy achieves a $32.2\%$ energy-efficiency gain over no-wait-direct and a $10.1\%$ gain over no-wait-assisted. Throughput decreases as $\tau_s$ grows, with the advantage diminishing for $\tau_s \geq 15$\,ms due to reduced time diversity.

\vspace{-0.2cm}
\section{Conclusion}\label{s:con}
\vspace{-0.2cm}
This study addresses the joint optimization of user scheduling and caching-aided access in hybrid satellite-terrestrial networks. We developed a novel analytical framework focusing on opportunistic scheduling through terrestrial stations probing and cache-aided delivery. Using optimal stopping theory, we derived an expectation function threshold based optimal stopping rule, implemented as a hybrid satellite-terrestrial joint probing and scheduling strategy. 
To overcome challenges like dynamic user requests, varying link conditions, and the coexistence of satellite and terrestrial caches, we propose an optimal strategy that maximizes system throughput with online complexity of $\mathcal{O}(1)$.
%We also introduced offline iterative algorithms for practical implementation.
Our strategy enhances system performance by leveraging caching gain, spatial diversity, and time diversity, adapting to network changes for improved throughput in hybrid satellite-terrestrial networks.


\vspace{-0.1cm}
\begin{thebibliography}{1}
\vspace{-0.1cm}
\bibliographystyle{IEEEtran}
%\bibitem{An2015}
%K. An, M. Lin and T. Liang, ``On the performance of multiuser hybrid satellite-terrestrial relay networks with opportunistic scheduling,'' \emph{IEEE Wireless Commun. Lett.}, vol. 19, no. 10, pp. 1722--1725, Aug. 2015.

\bibitem{Zhu2023}
L. Zhu, L. Bai, L. Zhou and J. Choi, ``Efficient user scheduling for uplink hybrid satellite-terrestrial communication,'' \emph{IEEE Trans. Wireless Commun.}, vol. 22, no. 3, pp. 1885--1899, Mar. 2023.

\bibitem{An2019}
K. An, Y. Li, X. Yan and T. Liang,``On the performance of cache-enabled hybrid satellite-terrestrial relay networks,'' \emph{IEEE Wireless Commun. Lett.}, vol. 8, no. 5, pp. 1506--1509, Oct. 2019.

\bibitem{Guo2024}
B. Guo et al., ``Enabling real-time computing and transmission services in large-scale LEO satellite networks,'' \emph{IEEE Trans. Veh. Technol.}, vol. 74, no. 8, pp. 12813--12828, Aug. 2025.

\bibitem{Zhao2022}
R. Zhao, Y. Ran, J. Luo and S. Chen, ``Towards coverage-aware cooperative video caching in LEO satellite networks,'' \emph{{IEEE} Global Commn. Conf. (GLOBECOM)}, Dec. 2022, pp. 1893--1898.
%\emph{GLOBECOM 2022 - 2022 IEEE Global Communications Conference}


\bibitem{Jiaran2023}
J. Zhang et al., ``Content-aware proportional caching for efficient data delivery over satellite network,'' \emph{{IEEE} Global Commn. Conf. (GLOBECOM)}, Dec. 2023, pp. 4890--4895.  

\bibitem{zhang2025IoTJ}
Z. Zhang et al., ``Optimizing energy-efficient cooperative {MAC} strategies for data collection in IoT networks with terrestrial and non-terrestrial relays," \emph{IEEE Internet Things J.}, vol. 17, pp. 35556--35576, Sep. 2025.

\bibitem{Han2022}
L. Han, W. -P. Zhu and M. Lin, ``Outage analysis of multi-relay NOMA-based hybrid satellite-terrestrial relay networks,'' \emph{IEEE Trans. Veh. Technol.}, vol. 71, no. 6, pp. 6469--6487, Jun. 2022.

%\bibitem{Zxuan_twc2024}
%X. Zhang, S. Sun, M. Tao, Q. Huang and X. Tang, ``Multi-satellite cooperative networks: joint hybrid beamforming and user scheduling design,'' \emph{IEEE Trans. Wireless Commun.}, vol. 23, no. 7, pp. 7938--7952, Jul. 2024.
\bibitem{Zhao2021}
B. Zhao et al, %G. Ren, X. Dong and H. Zhang, %, B. Zhao, M. Wang, Z. Xing, G. Ren and J. Su,  
``Distributed Q-learning based joint relay selection and access control scheme for IoT-oriented satellite terrestrial relay networks,'' \emph{IEEE Commun. Lett.}, vol. 25, pp. 1901--1905, Jun. 2021.

\bibitem{Yting2021}
L. Yan, X. Ding and G. Zhang, ``Dynamic channel allocation aided random access for SDN-enabled LEO satellite IoT,'' \emph{Journal of Commun. and Inform. Networks}, vol. 6, no. 2, pp. 134--141, Jun. 2021.

\bibitem{zhao2023wcl}
B. Zhao et al, %M. Wang, Z. Xing, G. Ren and J. Su,, %, B. Zhao, M. Wang, Z. Xing, G. Ren and J. Su,  
``Integrated sensing and communication aided dynamic resource allocation for random access in satellite terrestrial relay networks,'' \emph{IEEE Commun. Lett.}, vol. 27, no. 2, pp. 661--665, Feb. 2023.

\bibitem{zhang2024Journal}
Z. Zhang et al., ``Optimal cooperative MAC strategies for wireless VANETs with multiple roadside units," \emph{IEEE Trans. Veh. Technol.}, 
 vol. 1, pp. 877--893, Jan. 2025.
 
\bibitem{Ji2020}
Z. Ji, S. Wu, C. Jiang, D. Hu and W. Wang, ``Energy-efficient data offloading for multi-cell satellite-terrestrial networks,'' \emph{IEEE Commun. Lett.}, vol. 24, no. 10, pp. 2265--2269, Oct. 2020.


\bibitem{zhang2023distributedjournal}
Z. Zhang et al., ``Distributed {MAC} for {RIS}-assisted multiuser networks: {CSMA/CA} protocol design and statistical optimization," \emph{IEEE Trans. Mob. Comput.}, vol. 24, pp. 4698--4715, Jun. 2025.

\bibitem{Park2023}
J. Park et al., ``Random access protocol for massive internet of things connectivity in space–air–ground-integrated networks,'' \emph{IEEE Internet Things J.}, vol. 10, no. 23, pp. 20442 -- 20457, Jun. 2023.

%\bibitem{ZhangS2020}
%S. Zhang and J. Liu, ``Optimal probabilistic caching in heterogeneous IoT networks,'' \emph{IEEE Internet Things J.}, vol. 7, pp. 3404--3414, Apr. 2020.

%\bibitem{Minh2024}
%M. Nguyen et al., ``Real-time optimized clustering and caching for 6G satellite-UAV-terrestrial networks,'' \emph{{IEEE} Trans. Intell. Transp. Syst.}, vol. 25, no. 3, pp. 3009--3019, Mar. 2024.

%\bibitem{Sreng2012}
%S. Sreng et al., "Outage analysis of hybrid satellite-terrestrial cooperative network with best relay selection," Wireless Telecomm. Symposium 2012.

%\bibitem{Foddis2016}
%G. Foddis et al., ``Modeling RACH arrivals and collisions for human-type communication,'' \emph{IEEE Commun. Lett.}, vol. 20, pp. 1417--1420, Jul. 2016.

%\bibitem{LMS_model}
%S. Sreng et al.,  ``Outage analysis of hybrid satellite-terrestrial cooperative network with best relay selection,'' \emph{Wirel. Telecommun. Symp.}, Apr. 2012.

%\bibitem{Int_table}
%I. S. Gradshteyn amd I. M. Ryzhik, Table of Integrals, Series, and Products. New York: Academic Press, 2007.

\bibitem{Fenoy2017}
M. Feno, ``The invariant optimal sampling plan in a sequentially planned decision procedure,'' \emph{Sequential Analysis}, vol. 36, no. 2, pp. 194, 2017.


%\bibitem{xie2023}
%J. Xie et al., ``Optimal secure channel access in distributed cooperative networks with untrusted relay,'' \emph{IEEE Wireless Commun. Lett.}, vol. 12, no. 6, pp. 1091--1095, Mar. 2023.


%\bibitem{Schmitz_N_book}
%N. Schmitz,  \emph{Optimal sequentially planned decision procedures}, Lecture notes in statistics, New York: Springer-Verlag, 1993.

%\bibitem{Berts1999}
%D. Bertsekas, \emph{Nonlinear programming}, Athena Scientific, 1999.

%\bibitem{Zhu2022}
%X. Zhu, C. Jiang, L. Kuang and Z. Zhao, ``Cooperative multilayer edge caching in integrated satellite-terrestrial networks,'' \emph{IEEE Trans. Wireless Commun.}, vol. 21, no. 5, pp. 2924--2937, May. 2022.

\end{thebibliography}
\end{document}